\documentclass[preprint,prd,aps,showkeys,nofootinbib]{revtex4}
\usepackage{graphicx}
\usepackage{color}
\usepackage{CJK,upgreek,fancyhdr}
\usepackage{bm}
\usepackage{epstopdf}
\usepackage{multirow}

\begin{document}

\begin{CJK*}{GBK}{song}


\title{The origin of the 95 GeV excess in the flavor-dependent $U(1)_X$ model}

\author{Zhao-feng Ge$^{1,2}$\footnote{gezhaofeng@itp.ac.cn},Feng-Yan Niu$^{3,4,5}$,Jin-Lei Yang$^{3,4,5}$\footnote{jlyang@hbu.edu.cn}}

\affiliation{$^1$CAS Key Laboratory of Theoretical Physics, Institute of Theoretical Physics,
Chinese Academy of Sciences, Beijing 100190, China\\
$^2$School of Physical Sciences, University of Chinese
Academy of Sciences, Beijing 100049, China}

\affiliation{$^3$Department of Physics, Hebei University, Baoding, 071002, China\\
$^4$Key Laboratory of High-precision Computation and Application of Quantum Field Theory of Hebei Province, Baoding, 071002, China\\
$^5$Research Center for Computational Physics of Hebei Province, Baoding, 071002, China}

\begin{abstract}
This study investigates the excesses observed in the CMS diphoton and ditau data around $95\;{\rm GeV}$ within the framework of the flavor-dependent $U(1)_X$ model. The model introduces a singlet scalar to explain the nonzero neutrino masses. This newly introduced Higgs interacts directly with the quark sector, motivated by the aim to explain the flavor numbers of the fermion sector. Additionally, it undergoes mixing with the SM-like Higgs boson. The study suggests that designating this singlet Higgs state in this model as the lightest Higgs boson holds great potential for explaining the excesses around $95\;{\rm GeV}$. In the calculations, we maintained the masses of the lightest and next-to-lightest Higgs bosons at around $95\;{\rm GeV}$ and $125\;{\rm GeV}$ respectively. It was found that the theoretical predictions on the signal strengthes $\mu(h_{95})_{\gamma\gamma}$, $\mu(h_{95})_{\tau\tau}$ in the flavor-dependent $U(1)_X$ model can be fitted well to the excesses observed at CMS.

\end{abstract}


\maketitle

\section{Introduction\label{sec1}}

The discovery of a $125\;{\rm GeV}$ Higgs boson at the Large Hadron Collider (LHC) in 2012~\cite{CMS:2012qbp,ATLAS:2012yve} stands as one of the most remarkable achievements in theoretical physics. Its properties align well with the predictions of the standard model (SM)~\cite{CMS:2022dwd,ATLAS:2022vkf}, suggesting the observation of all fundamental particles anticipated by the SM.

The current focus of the LHC program is to determine whether the detected Higgs boson is the sole fundamental scalar particle or part of a new physics (NP) theory with an extended Higgs sector. As of now, no new scalars have been discovered at the LHC. However, there have been several intriguing excesses observed in searches for light Higgs bosons below 125 GeV, accompanied by increasing precision in the measurements of the Higgs couplings to fermions and gauge bosons.

The results derived from both the CMS Run 1 and the first year of CMS Run 2 data for Higgs-boson searches in the diphoton final state show a $2.8\sigma$ local excess at a mass about $95\;{\rm GeV}$~\cite{CMS:2015ocq,CMS:2018cyk}, which is compatible with the latest ATLAS result~\cite{ATLAS:2023CA} based on the previously reported result utilizing $80\;{\rm fb}^{-1}$~\cite{ATLAS:2018xad} in the diphoton searches
\begin{eqnarray}
&&\mu(\Phi_{95})_{\gamma\gamma}=\frac{\sigma(gg\rightarrow \Phi^{\rm NP}_{95}){\rm BR}(\Phi^{\rm NP}_{95}\rightarrow \gamma\gamma)}{\sigma(gg\rightarrow h^{\rm SM}_{95}){\rm BR}(h^{\rm SM}_{95}\rightarrow \gamma\gamma)}=0.18\pm0.10.
\end{eqnarray}
Utilizing the full Run 2 dataset, CMS published the results for the Higgs boson searches in the $\tau^+\tau^-$ channel , indicating a local significance of $3.1\sigma$ at a mass value around $95\;{\rm GeV}$~\cite{CMS:2022goy}
\begin{eqnarray}
&&\mu(\Phi_{95})_{\tau\tau}=\frac{\sigma(gg\rightarrow \Phi^{\rm NP}_{95}){\rm BR}(\Phi^{\rm NP}_{95}\rightarrow \tau\tau)}{\sigma(gg\rightarrow h^{\rm SM}_{95}){\rm BR}(h^{\rm SM}_{95}\rightarrow \tau\tau)}=1.2\pm0.5.\label{eq2}
\end{eqnarray}
Recently, a new analysis of Higgs-boson searches in the diphoton final state by CMS has further confirmed the excess at approximately $95\;{\rm GeV}$. The results are reported in~\cite{CMS:2023}.
\begin{eqnarray}
&&\mu(\Phi_{95})_{\gamma\gamma}=0.33_{-0.12}^{+0.19}.\label{eq3}
\end{eqnarray}

Theoretically, there are numerous discussions on the excesses in NP models. The analysis carried out in Refs.~\cite{Moretti:2006sv,Ellwanger:2010nf,
Cao:2011pg,AlbornozVasquez:2012foo,Ellwanger:2012ke,Boudjema:2012in,Schmidt-Hoberg:2012dba,Badziak:2013bda,Badziak:2013gla,Barbieri:2013nka,Fan:2013gjf,
Potter:2015wsa,Ellwanger:2015uaz,Cao:2016uwt,Cao:2019ofo} indicates that the diphoton rate may be several times larger than its SM prediction for the same scalar mass in the next-to-minimal supersymmetric standard model (NMSSM). In the Two-Higgs doublet model (2HDM) with an additional real singlet (N2HDM), the possibilities of explaining the observed excesses are studied in Refs.~\cite{Biekotter:2019kde,Biekotter:2021ovi,Biekotter:2021qbc,Heinemeyer:2021msz,Biekotter:2022jyr,Biekotter:2022abc,Biekotter:2023jld,
Azevedo:2023zkg,Aguilar-Saavedra:2023vpd}. The authors of Ref.~\cite{Sachdeva:2019hvk} explore the viability of the radion mixed Higgs to be the 125 GeV boson along with the presence of a light radion which can account for the CMS diphoton excess well in the Higgs-radion mixing model. Considering the one-loop corrections to the neutral scalar masses of the $\mu\nu$SSM, the authors of Refs.~\cite{Biekotter:2017xmf,Biekotter:2019gtq} demonstrate how the $\mu\nu$SSM can simultaneously accommodate two excesses measured at the Large Electron-Positron~\cite{LEP:2003ing} and LHC at the $1\sigma$ level. Based on the analysis in Ref.~\cite{Ashanujjaman:2023etj}, extending the scalar sector with a $SU(2)_L$ triplet with hypercharge $Y=0$ can well provide the origin of the $95\;{\rm GeV}$ excesses. Whether certain model realizations could simultaneously explain the two excesses while being in agreement with all other Higgs-boson related limits and measurements were reviewed in Refs.~\cite{Azatov:2012bz,Heinemeyer:2018wzl}.

In the flavor-dependent $U(1)_X$ model, the new $U(1)_X$ charge is related to the baryon and lepton number, and the origin of the number of observed fermion families can be well explained by imposing a relation involved by the color number $3$~\cite{VanLoi:2023utt}. Because of the $Z_2$ conservation, the model can contain several scenarios for single-component dark matter. In addition, the nonzero neutrino masses can be obtained elegantly by type-I see-saw mechanism by the new introduced singlet scalar and the nonzero $U(1)_X$ charges of neutrinos. As a result, the new introduced CP-even scalar may be assigned to account for the excesses at about $95\;{\rm GeV}$. Based on the characters of the new introduced CP-even scalar in the flavor-dependent $U(1)_X$ model, we focus on investigating the diphoton and ditau excesses at about $95\;{\rm GeV}$ in this work\footnote{As analyzed in Refs.~\cite{CMS:2022arx,Iguro:2022dok}, the light CP-even Higgs does suffers strict constraints from the LHC search for the top-quark associated production of the SM Higgs boson that decays into $\tau\bar\tau$, and the possibility to explain the ditau excess by the CP-even scalar is excluded according the analysis in Ref.~\cite{Iguro:2022dok}. However, they consider the $1\sigma$ range of $\mu(\Phi_{95})_{\tau\tau}$ in the analysis, it is found that this constraint can be relaxed if the $2\sigma$ range of $\mu(\Phi_{95})_{\tau\tau}$ is considered. And the coupling properties of $125\;{\rm GeV}$ Higgs in the flavor-dependent $U(1)_X$ model are different from the ones of $125\;{\rm GeV}$ Higgs in the SM, so the 125 GeV signal strength constraints are also taken into account in the analysis.}.

The paper is organized as follows. The scalar sector and fermion masses terms of the flavor-dependent $U(1)_X$ model are given in Sec.~\ref{sec2}. The numerical results are present and analyzed in Sec.~\ref{sec3}. Finally, a summary is made in Sec.~\ref{sec4}.

\section{The $U(1)_X$ model and its Higgs sector\label{sec2}}

The flavor-dependent $U(1)_X$ model~\cite{VanLoi:2023utt} is one of the simplest extension of the SM that introduces a flavor-dependent additional gauge group $U(1)_X$. The gauge group of the model is $SU(3)_C\otimes SU(2)_L \otimes U(1)_Y\otimes U(1)_X$ in the model, and local $U(1)_X$ gauge group is not family universal. The new charge $X$, combines lepton ($L$) and baryon ($B$) numbers, are defined as
\begin{eqnarray}
X=xB+yL,
\end{eqnarray}
and $X$ depends on family. For each flavor $i$, $X=x_ib +y_iL$, where $x_i$ and $y_i$ are functions of $i$ for $i=1,2,...,N_f$, while $B$ and $L$ denotes the total baryon or lepton numbers, respectively. The SM fermions transform under the gauge symmetry as
\begin{eqnarray}
&&l_{iL}=\left(\begin{array}{c}\nu_{iL}\\ e_{iL}\end{array}\right)\sim(1,2,-1/2,y_i),\;q_{iL}=\left(\begin{array}{c}u_{iL}\\ d_{iL}\end{array}\right)\sim(3,2,1/6,x_i/3),\nonumber\\
&&e_{iR}\sim(1,1,-1,y_i),\;u_{iR}\sim(3,1,2/3,x_i/3),\;d_{iR}\sim(3,1,-1/3,x_i/3),
\end{eqnarray}
where $x_i$ are family dependent, and $y_i$ do not. The charges of all fields in this model are presented in Table~\ref{table:1}, where $a=1,2,3$, $\alpha=1,2$, and $z$ is arbitrarily nonzero.
\begin{table}[htb]
\begin{center}
\begin{tabular}{|c|c|c|c|c|c|}
\hline   Multiplets & $SU(3)_C$ & $SU(2)_L$& $U(1)_Y$& $U(1)_X$&$Z_2$ \\
\hline   $l_{aL}=(\nu_{aL},e_{aL})^T$ & 1 & 2 & $-\frac{1}{2}$ & z & +\\
 $\nu_{aR}$ & 1 & 1 & 0 & z & +\\
 $e_{aR}$ & 1 & 1 & -1 & z & +\\
 $q_{\alpha L}=(u_{\alpha L},d_{\alpha L})^T$ & 3 & 2 & $\frac{1}{6}$ & z & +\\
 $u_{\alpha R}$ & 3 & 1 & $\frac{2}{3}$ & -z & +\\
 $d_{\alpha R}$ & 3 & 1 & $-\frac{1}{3}$ & -z & +\\
 $q_{3 L}=(u_{3 L},d_{3 L})^T$ & 3 & 2 & $\frac{1}{6}$ & z & +\\
 $u_{3 R}$ & 3 & 1 & $\frac{2}{3}$ & z & +\\
 $d_{3 R}$ & 3 & 1 & $-\frac{1}{3}$ & z & +\\
 $\Phi=(\Phi_1^+,\Phi_2^0)^T$ & 1 & 2 & $\frac{1}{2}$ & 0 & +\\
 $\chi$ & 1 & 1 & 0 & -2z & +\\
 $\xi$ & 1 & 1 & 0 & 2z & -\\
 $\eta$ & 1 & 1 & 0 & z & -\\
\hline
\end{tabular}
\end{center}
\caption{Multiplets in the flavor-dependent $U(1)_X$ model~\cite{VanLoi:2023utt}.}
\label{table:1}
\end{table}

This model can explain the origin of the observed number of fermion families and potentially offer solutions for both neutrino mass and dark matter, which differs from the traditional $U(1)_{B-L}$ extension. The total scalar potential in this model is given by~\cite{VanLoi:2023utt}
\begin{eqnarray}
&&V=\mu_1^2\Phi^\dagger\Phi+\mu_2^2\chi^*\chi+\mu_3^2\eta^*\eta+\lambda_1(\Phi^\dagger\Phi)^2
+\lambda_2(\chi^*\chi)^2+\lambda_3(\eta^*\eta)^2 +\nonumber\\
&&\qquad\lambda_4(\Phi^\dagger\Phi)(\chi^*\chi)
+\lambda_5(\Phi^\dagger\Phi)(\eta^*\eta)+\lambda_6(\chi^*\chi)(\eta^*\eta)
\label{scalarpotential}
\end{eqnarray}
where $\lambda$'s are dimensionless,and $\mu$'s have mass dimension. To ensure the stability of the scalar potential, the free parameters $\lambda_1,\;\lambda_2,\;\lambda_3,\;\mu_1,\;\mu_2,\;\mu_3$ should satisfy
\begin{eqnarray}
\lambda_{1,2,3}>0,\;\mu_{1,2}^2<0,\;\mu_3^2>0.
\end{eqnarray}
The local gauge symmetry $SU(2)_L \otimes U(1)_Y\otimes U(1)_{X}$ breaks down to the electromagnetic symmetry $U(1)_{\rm em}$ as the scalar fields receive nonzero vacuum expectation values (VEV):
\begin{eqnarray}
&&\Phi=\left(\begin{array}{c}\Phi^+_1\\ \frac{1}{\sqrt{2}}(v+S_1+iA_1)\end{array}\right),\chi=\frac{1}{\sqrt{2}}(\Lambda+S_2+iA_2),
\eta=\frac{1}{\sqrt{2}}(S_3+iA_3).
\end{eqnarray}
Substituting them into the scalar potential Eq.~(\ref{scalarpotential}), the tadpole equations give
\begin{eqnarray}
&&\mu_1^2=-\frac{1}{2}(2\lambda_1\;v^2+\lambda_4\;\Lambda^2),\;\;\;
\mu_2^2=-\frac{1}{2}(2\lambda_2\;\Lambda^2+\lambda_4\;v^2).
\label{potentialminimum}
\end{eqnarray}
Due to $Z_2$ conservation, the dark scalars $S_3$ and $A_3$ do not mix with the other scalars and are degenerate in mass. On the basis $(S_1,\;S_2,\;S_3)$, the squared mass matrix of Higgs can be written as
\begin{eqnarray}
&&M_{h}^2=\left(\begin{array}{*{20}{c}}
2\lambda_1 v^2 & \lambda_4 v \Lambda & 0 \\ [6pt]
\lambda_4 v \Lambda & 2\lambda_2 \Lambda^2 & 0 \\ [6pt]
0 & 0 & \frac{1}{2}(\lambda_5 v^2+\lambda_6 \Lambda^2+2\mu_3^2) \\ [6pt]
\end{array}\right),\label{eq8}
\end{eqnarray}
where $S_1$ and $S_2$ are mixed by the term $\lambda_4 v \Lambda$. The squared mass matrix $M_h^2$ in Eq.~(\ref{eq8}) can be diagonalized by the unitary matrix $Z_h$ as
\begin{eqnarray}
&&{\rm diag}(m_{h_1}^2,m_{h_2}^2,m_{h_3}^2)=Z_h\cdot M_h^2\cdot Z_h^\dagger,
\end{eqnarray}
where $m_{h_1},m_{h_2},m_{h_3}$ are the physical Higgs masses. And under the assumption $v/\Lambda\ll1$, $\lambda_2\ll1$ and $\lambda_4\ll1$, we can obtain
\begin{eqnarray}
&&m_{h_1}^2\approx 2\lambda_2\Lambda^2+\frac{\lambda_4^2}{2\lambda_2} v^2,\;m_{h_2}^2\approx 2\lambda_1 v^2,\;m_{h_3}^2=\frac{1}{2}(\lambda_5 v^2+\lambda_6 \Lambda^2+2\mu_3^2).\label{amh}
\end{eqnarray}
$Z_h$ can be parameterized as
\begin{eqnarray}
&&Z_{h}=\left(\begin{array}{*{20}{c}}
-\sin\theta&\cos\theta&0 \\ [6pt]
\cos\theta&\sin\theta&0 \\ [6pt]
0&0&1 \\ [6pt]
\end{array}\right),\label{Zh}
\end{eqnarray}
where the mixing angle $\cos^2\theta=\frac{1}{2}(\frac{1}{\sqrt{1+A^2}}+1)$ with $A=-\frac{\lambda_4\;v\;\Lambda}{\lambda_1\;v^2-\lambda_2\;\Lambda^2}$.

After spontaneous symmetry breaking, the fermion mass terms are given by the Yukawa couplings
\begin{eqnarray}
&&\mathcal{L}\sim h_{ab}^e\bar{l}_{aL}\Phi e_{bR}+h_{ab}^v\bar{l}_{aL}\tilde{\Phi}\nu_{bR}+\frac{1}{2}f_{ab}^\nu \bar{\nu}_{aR}^c \nu_{bR}\chi+h_{\alpha \beta}^d\bar{q}_{\alpha L}\Phi d_{\beta R}+h_{\alpha \beta}^u\bar{q}_{\alpha L}\tilde{\Phi}u_{\beta R}\nonumber\\
&&\qquad +h_{33}^d \bar{q}_{3L}\Phi d_{3R}+h_{33}^u \bar{q}_{3L}\tilde{\Phi}u_{3R}+\frac{h_{\alpha 3}^d}{M}\bar{q}_{\alpha L}\Phi\chi d_{3R}+\frac{h_{3\alpha}^u}{M}\bar{q}_{3L}\tilde{\Phi}\chi^*u_{\alpha R}\nonumber\\
&&\qquad +\frac{h_{3\alpha}^d}{M}\bar{q}_{3L}\Phi\chi^*d_{\alpha R}
+\frac{h_{\alpha 3}^u}{M}\bar{q}_{\alpha L}\tilde{\Phi}\chi u_{3R}+ y_{a}\bar{\xi}_L \eta \nu_{aR}-m_{\xi}\bar{\xi}_L\xi_R+h.c.,
\end{eqnarray}
where $a,b=1,2,3$, $\alpha,\beta=1,2$. The fermion-antifermion-higgs couplings on the interactional basis can be written as
\begin{eqnarray}
&&\mathcal{L}\sim \frac{1}{\sqrt{2}}h_{ab}^e \bar{e}_{aL} S_1 e_{bR}+\frac{1}{\sqrt{2}}h_{ab}^\nu \bar{\nu}_{aL}S_1 \nu_{bR}+\frac{1}{\sqrt{2}}h_{\alpha \beta}^d \bar{d}_{\alpha L}S_1 d_{\beta R}+\nonumber\\
&&\qquad \frac{1}{\sqrt{2}}h_{\alpha \beta}^u \bar{u}_{\alpha L}S_1 u_{\beta R}+\frac{1}{\sqrt{2}}h_{33}^d \bar{d}_{3L}S_1 d_{3R}+\frac{1}{\sqrt{2}}h_{33}^u \bar{u}_{3L}S_1 u_{3R}+\nonumber\\
&&\qquad \frac{h_{\alpha 3}^d}{2M}(v \bar d_{\alpha L}S_2 d_{3R}+\Lambda \bar d_{\alpha L}S_1 d_{3R})+\frac{h_{3\alpha}^u}{2M}(v \bar u_{3 L}S_2 u_{\alpha R}+\Lambda \bar u_{3 L}S_1 u_{\alpha R})+\nonumber\\
&&\qquad \frac{h_{3\alpha}^d}{2M}(v \bar d_{3 L}S_2 d_{\alpha R}+\Lambda \bar d_{3 L}S_1 d_{\alpha R})
+\frac{h_{\alpha 3}^u}{2M}(v \bar u_{\alpha L}S_2 u_{3R}+\Lambda \bar u_{\alpha L}S_1 u_{3R}).
\end{eqnarray}
It is worth noting the presence of additional terms compared to the quark mass matrices in the SM, such as $\frac{h_{3\alpha}^d}{2M}(v \bar d_{3 L}S_2 d_{\alpha R}+\Lambda \bar d_{3 L}S_1 d_{\alpha R})$ in the down-quark sector, $\frac{h_{3\alpha}^u}{2M}(v \bar u_{3 L}S_2 u_{\alpha R}+\Lambda \bar u_{3 L}S_1 u_{\alpha R})$ in the up-quark sector and etc. Under the minimal flavor violation assumption~\cite{Chivukula:1987py,Hall:1990ac,Buras:2000dm,DAmbrosio:2002vsn,Isidori:2010kg} which can release the model from the experimental constraints on the processes mediated by flavor-changing neutral currents (FCNCs), the fermion-antifermion-higgs couplings on the physical basis $h_1\bar f_i^m f_i^m,\;(q=u,d,e)$ can be written as
\begin{eqnarray}
&&\mathcal{L_I}=h_1(\bar{d}_1^m,\bar{d}_2^m,\bar{d}_3^m)\left(\begin{array}{*{20}{c}}
m_d/v & 0 & 0 \\ [6pt]
0 & m_s/v & 0 \\ [6pt]
0 & 0 & m_b/v \\ [6pt]
\end{array}\right)\left(\begin{array}{*{20}{c}}
d_1^m \\ [6pt]
d_2^m \\ [6pt]
d_3^m \\ [6pt]
\end{array}\right)(Z_{11}^h+Z_{21}^h \kappa_1)+\nonumber\\
&&\qquad\; h_1(\bar{u}_1^m,\bar{u}_2^m,\bar{u}_3^m)\left(\begin{array}{*{20}{c}}
m_u/v & 0 & 0 \\ [6pt]
0 & m_c/v & 0 \\ [6pt]
0 & 0 & m_t/v \\ [6pt]
\end{array}\right)\left(\begin{array}{*{20}{c}}
u_1^m \\ [6pt]
u_2^m \\ [6pt]
u_3^m \\ [6pt]
\end{array}\right)(Z_{11}^h+Z_{21}^h \kappa_2)+\nonumber\\
&&\qquad\; h_1(\bar{e}_1^m,\bar{e}_2^m,\bar{e}_3^m)\left(\begin{array}{*{20}{c}}
m_e/v & 0 & 0 \\ [6pt]
0 & m_\mu/v & 0 \\ [6pt]
0 & 0 & m_\tau/v \\ [6pt]
\end{array}\right)\left(\begin{array}{*{20}{c}}
e_1^m \\ [6pt]
e_2^m \\ [6pt]
e_3^m \\ [6pt]
\end{array}\right)(Z_{11}^h).\label{eqak}
\end{eqnarray}
Here, the contributions from the additional terms in the quark sector are absorbed into the constants $\kappa_1$ and $\kappa_2$ with
\begin{eqnarray}
&&\kappa_1 m_{d_i}=\Big[Z_L^{d\dagger}\left(\begin{array}{*{20}{c}}
0 & 0 & -h^d_{13}\frac{v\Lambda}{2M} \\ [6pt]
0 & 0 & -h^d_{23}\frac{v\Lambda}{2M} \\ [6pt]
-h^d_{31}\frac{v\Lambda}{2M} & -h^d_{32}\frac{v\Lambda}{2M} & 0 \\ [6pt]
\end{array}\right)Z_R^d\Big](i,i),\nonumber\\
&&\kappa_2 m_{u_i}=\Big[Z_L^{u\dagger}\left(\begin{array}{*{20}{c}}
0 & 0 & -h^u_{13}\frac{v\Lambda}{2M} \\ [6pt]
0 & 0 & -h^u_{23}\frac{v\Lambda}{2M} \\ [6pt]
-h^u_{31}\frac{v\Lambda}{2M} & -h^u_{32}\frac{v\Lambda}{2M} & 0 \\ [6pt]
\end{array}\right)Z_R^u\Big](i,i),
\end{eqnarray}
where $m_{q_i}$ denotes the $i-th$ quark $q$ mass, $Z_L^q$ and $Z_R^q$ are the unitary matrices which diagonalize the mass matrix of quark $q$, and $(i,i)$ denotes the $i-th$ diagonal element of matrix. It is obvious that $\kappa_1,\;\kappa_2$ depend on the parameters $\Lambda,\;M$ and the Yukawa coupling constant complicatedly. For simplicity, as shown in Eq.~(\ref{eqak}), we take the diagonal elements of the matrices defined as $\kappa_1 m_{d_i}$, $\kappa_2 m_{u_i}$ as inputs to carry out the analysis.

As defined in the introduction sector, the diphoton and ditau signal strength are
\begin{eqnarray}
&&\mu(h_{95})_{\gamma\gamma}=\frac{\sigma(gg\rightarrow h^{\rm NP}_{95}){\rm BR}(h^{\rm NP}_{95}\rightarrow \gamma\gamma)}{\sigma(gg\rightarrow h^{\rm SM}_{95}){\rm BR}(h^{\rm SM}_{95}\rightarrow \gamma\gamma)},\nonumber\\
&&\mu(h_{95})_{\tau\tau}=\frac{\sigma(gg\rightarrow h^{\rm NP}_{95}){\rm BR}(h^{\rm NP}_{95}\rightarrow \tau\tau)}{\sigma(gg\rightarrow h^{\rm SM}_{95}){\rm BR}(h^{\rm SM}_{95}\rightarrow \tau\tau)},
\end{eqnarray}
where~\cite{Djouadi:1997yw}
\begin{eqnarray}
&&\Gamma^{\rm SM}_{\rm tot,95}\approx 0.00259\;{\rm GeV},\nonumber\\
&&{\rm BR}(h^{\rm SM}_{95}\rightarrow \gamma\gamma)\approx 1.4\times10^{-3},\;\;{\rm BR}(h^{\rm SM}_{95}\rightarrow \tau\tau)\approx0.082,
\end{eqnarray}
and the top quark loop is considered in calculating the production cross section of the $95$~GeV Higgs at the LHC for simplicity. The contributions from NP can be written as~\cite{Cao:2016uwt}
\begin{eqnarray}
&&\Gamma^{\rm NP}_{{\rm tot},95}\approx C_{h_{95}dd}^2\Gamma^{\rm SM}_{b\bar b,95}+C_{h_{95}ee}^2\Gamma^{\rm SM}_{\tau\bar \tau,95}+C_{h_{95}uu}^2(\Gamma^{\rm SM}_{c\bar c,95}+\Gamma^{\rm SM}_{gg,95})\nonumber\\
&&{\rm BR}(h^{\rm NP}_{95}\rightarrow \gamma\gamma)\approx C_{h_{95}uu}^2 {\rm BR}(h^{\rm SM}_{95}\rightarrow \gamma\gamma)\frac{\Gamma^{\rm SM}_{\rm tot,95}}{\Gamma^{\rm NP}_{\rm tot,95}},\nonumber\\
&&{\rm BR}(h^{\rm NP}_{95}\rightarrow \tau\bar \tau)\approx C_{h_{95}ee}^2 {\rm BR}(h^{\rm SM}_{95}\rightarrow \tau\bar \tau)\frac{\Gamma^{\rm SM}_{\rm tot,95}}{\Gamma^{\rm NP}_{\rm tot,95}},
\end{eqnarray}
where the coefficients $C_{h_{95}uu},\;C_{h_{95}dd},\;C_{h_{95}ee}$ are the normalized couplings of $95\;{\rm GeV}$ Higgs in NP models with SM particles (in units of the corresponding SM couplings). In the conventional $U(1)_X$ model, we have
\begin{eqnarray}
&&C_{h_{95}uu}=C_{h_1uu}=Z_{h,11}+  Z_{h,21} \kappa_1,\;\;C_{h_{95}dd}=C_{h_1 dd}=Z_{h,11}+ Z_{h,21}\kappa_2,\nonumber\\
&&C_{h_{95}ee}=C_{h_1 dd}=Z_{h,11}.
\end{eqnarray}

As shown above, the extra scalar singlet (which is designed to be the $95$ GeV scalar state) couples to the SM quarks at the tree level, and the corresponding effects are collected to the newly defined parameters $\kappa_1,\kappa_2$. Hence, there are two sources to accommodate the excesses in the diphoton and ditau channels in the model. The first one comes from the tree level couplings between the extra scalar singlet with the SM quarks, and the second one comes from the mixing of the $95$ GeV scalar state with the $125$ GeV one. It indicates that the $125$ GeV Higgs boson signal strength measurements should to be considered in the analysis. Generally, the signal strengths for the $125$ GeV Higgs decay channels can be quantified as~\cite{Arbey}
\begin{eqnarray}
&&\mu(h_{125})_{\gamma\gamma,VV^*}^{\rm ggF}=\frac{\sigma(gg\rightarrow h^{\rm NP}_{125}){\rm BR}(h^{\rm
NP}_{125}\rightarrow \gamma\gamma,VV^*)}{\sigma(gg\rightarrow h^{\rm SM}_{125}){\rm BR}(h^{\rm SM}_{125}\rightarrow \gamma\gamma,VV^*)},(V=W,\;Z)\nonumber\\
&&\mu(h_{125})_{ff}^{\rm VBF}=\frac{\sigma(VV^*\rightarrow h^{\rm NP}_{125}){\rm BR}(h^{\rm NP}_{125}\rightarrow f\bar{f})}{\sigma(VV^*\rightarrow h^{\rm SM}_{125}){\rm BR}(h^{\rm SM}_{125}\rightarrow f\bar{f})},(f=b,\;\tau,\;c),
\end{eqnarray}
where ggF and VBF stand for gluon-gluon fusion and vector boson fusion respectively. The corresponding SM decay width and fractions can be found in ref~\cite{PDG}.
The contributions from NP can be written as
\begin{eqnarray}
&&\Gamma^{\rm NP}_{{\rm tot},125}\approx C_{h_{125}dd}^2\Gamma^{\rm SM}_{b\bar b,125}+C_{h_{125}uu}^2\Gamma^{\rm SM}_{c\bar c,125}+C_{h_{125}ee}^2\Gamma^{\rm SM}_{\tau\bar \tau,125}+C_{h_{125}VV}^2(\Gamma^{\rm SM}_{WW^*,125}+\Gamma^{\rm SM}_{ZZ^*,125}),\nonumber\\
&&{\rm BR}(h^{\rm NP}_{125}\rightarrow \gamma\gamma)\approx C_{h_{125}uu}^2 {\rm BR}(h^{\rm SM}_{125}\rightarrow \gamma\gamma)\frac{\Gamma^{\rm SM}_{\rm tot,125}}{\Gamma^{\rm NP}_{\rm tot,125}},\nonumber\\
&&{\rm BR}(h^{\rm NP}_{125}\rightarrow \tau\bar \tau)\approx C_{h_{125}ee}^2 {\rm BR}(h^{\rm SM}_{125}\rightarrow \tau\bar \tau)\frac{\Gamma^{\rm SM}_{\rm tot,125}}{\Gamma^{\rm NP}_{\rm tot,125}},\nonumber\\
&&{\rm BR}(h^{\rm NP}_{125}\rightarrow b\bar b)\approx C_{h_{125}dd}^2 {\rm BR}(h^{\rm SM}_{125}\rightarrow \tau\bar \tau)\frac{\Gamma^{\rm SM}_{\rm tot,125}}{\Gamma^{\rm NP}_{\rm tot,125}},\nonumber\\
&&{\rm BR}(h^{\rm NP}_{125}\rightarrow VV^*)\approx C_{h_{125}VV}^2 {\rm BR}(h^{\rm SM}_{125}\rightarrow \tau\bar \tau)\frac{\Gamma^{\rm SM}_{\rm tot,125}}{\Gamma^{\rm NP}_{\rm tot,125}},
\end{eqnarray}
where the coefficients $C_{h_{125}uu}$, $C_{h_{125}dd}$, $C_{h_{125}ee}$ and $C_{h_{125}VV}$ are the normalized couplings of 125 GeV Higgs
in NP models with SM particles (in units of the corresponding SM couplings). In the conventional $U(1)_X$ model, we have
\begin{eqnarray}
&&C_{h_{125}uu}=C_{h_2uu}=Z_{h,12}+  Z_{h,22} \kappa_1,\;\;C_{h_{125}dd}=C_{h_2 dd}=Z_{h,12}+ Z_{h,22}\kappa_2,\nonumber\\
&&C_{h_{125}ee}=C_{h_2 dd}=Z_{h,12},\;\;C_{h_{125}VV}=C_{h_2 VV}=Z_{h,12}.
\end{eqnarray}

\section{Numerical results\label{sec3}}

\begin{figure}
\setlength{\unitlength}{1mm}
\centering
\includegraphics[width=2.1in]{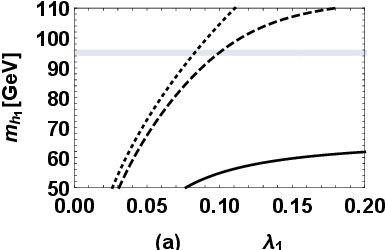}
\vspace{0.5cm}
\includegraphics[width=2.1in]{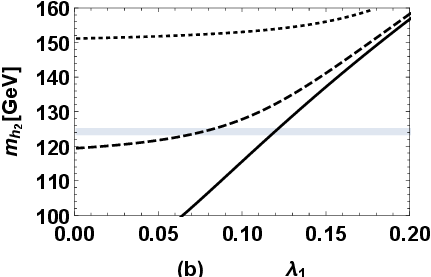}
\vspace{0.5cm}
\includegraphics[width=2.1in]{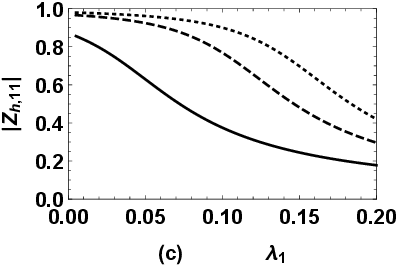}
\vspace{0cm}
\caption[]{Taking $\Lambda=15\;{\rm TeV}$, $\lambda_4=0.001$, the mass of the lightest Higgs $m_{h_1}$ versus $\lambda_1$ is plotted in (a), the mass of the next-to-lightest Higgs versus $\lambda_1$ is plotted in (b), and the Higgs mixing parameter $|Z_{h,11}|$ versus $\lambda_4$ is plotted in (c). The solid, dashed, dotted lines denote the results for $\lambda_2=0.00001,\;0.00003,\;0.00005$ respectively, the gray areas denote the range $94\;{\rm GeV}<m_{h_1}<96\;{\rm GeV}$ for (a) and $124\;{\rm GeV}<m_{h_2}<126\;{\rm GeV}$ for (b).}
\label{Fig1}
\end{figure}

Assuming the mass of the lightest Higgs boson to be approximately $95\;{\rm GeV}$ and the next-to-lightest Higgs boson corresponds to the measured SM-like Higgs~\cite{PDG} with the mass given by
\begin{eqnarray}
m_h = 125.09\pm0.24\;{\rm GeV},
\end{eqnarray}
we present the numerical results for the signal strengths of diphoton and ditau processes in this section. The outcomes of the process are primarily influenced by parameters $\lambda_1$, $\lambda_2$, $\lambda_4$, $\Lambda$, $\kappa_1$ and $\kappa_2$. Here, $\lambda_1$, $\lambda_2$, $\lambda_4$ are constants in the scalar sectors. $\Lambda$ is the vacuum expectation value scalar $S_2$, and it is constrained by data from FCNC and particle colliders, necessitating it to be larger than $14.14\;{\rm TeV}$~\cite{VanLoi:2023utt} which can be released under the minimal flavor violation assumption above.

Firstly, we investigate the masses of the lightest and next-to-lightest Higgs bosons within the model. Taking $\Lambda=15\;{\rm TeV}$ and $\lambda_4=0.001$, we plot the mass of the lightest Higgs $m_{h_1}$ versus $\lambda_1$ in Fig.~\ref{Fig1} (a), the mass of the next-to-lightest Higgs $m_{h_2}$ versus $\lambda_1$ in Fig.~\ref{Fig1} (b), and the Higgs mixing parameter $|Z_{h,11}|$ versus $\lambda_4$ in Fig.~\ref{Fig1} (c), where the solid, dashed, dotted lines denote the results for $\lambda_2=0.00001,\;0.00003,\;0.00005$ respectively. The gray areas denote the range $94\;{\rm GeV}<m_{h_1}<96\;{\rm GeV}$ for Fig.~\ref{Fig1} (a) and $124\;{\rm GeV}<m_{h_2}<126\;{\rm GeV}$ for Fig.~\ref{Fig1} (b). It can be observed from the plot that $\lambda_1$ and $\lambda_2$ significantly influence the masses of the two light Higgs bosons, and both $m_{h_1}$ and $m_{h_2}$ increase as $\lambda_1$ increases. Notably, both the $95\;{\rm GeV}$ Higgs and the SM-like Higgs with a mass of $125\;{\rm GeV}$ are attainable within the model. Fig.~\ref{Fig1} (a) shows that $\lambda_2$ is constrained around $0.00003$ with $\lambda_1$ slightly less than $0.1$ for $m_{h_1}\approx 95\;{\rm GeV}$ and $m_{h_2}\approx 125\;{\rm GeV}$ in our chosen parameter space. $\lambda_1$, $\lambda_2$ and $\lambda_4$ collectively determine the two light Higgs masses.

\begin{figure}
\setlength{\unitlength}{1mm}
\centering
\includegraphics[width=2.1in]{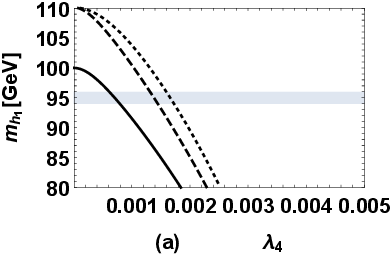}
\vspace{0.5cm}
\includegraphics[width=2.1in]{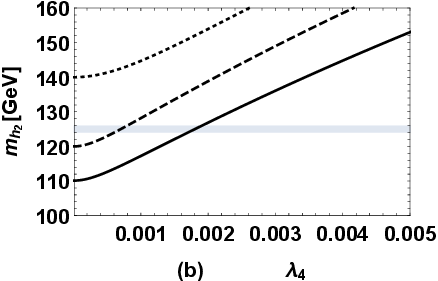}
\vspace{0.5cm}
\includegraphics[width=2.1in]{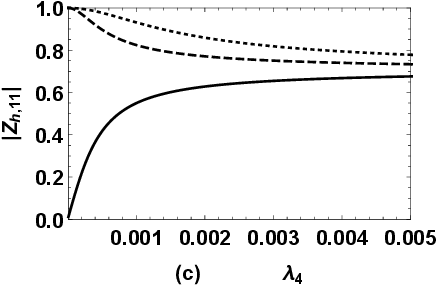}
\vspace{0cm}
\caption[]{Taking $\lambda_1=0.1$, $\lambda_2=0.00005$, the mass of the lightest Higgs $m_{h_1}$ versus $\lambda_4$ is plotted in (a), the mass of the next-to-lightest Higgs $m_{h_2}$ versus $\lambda_4$ is plotted in (b), and the Higgs mixing parameter $|Z_{h,11}|$ versus $\lambda_4$ is plotted in (c). The solid, dashed, dotted lines denote the results for $\Lambda=10,\;12,\;14{\rm TeV}$ respectively, the gray areas denote the range $94\;{\rm GeV}<m_{h_1}<96\;{\rm GeV}$ for (a) and $124\;{\rm GeV}<m_{h_2}<126\;{\rm GeV}$ for (b).}
\label{Fig2}
\end{figure}

Setting $\lambda_1=0.1$, $\lambda_2=0.00005$, we plot $m_{h_1}$, $m_{h_2}$ and the Higgs mixing parameter $|Z_{h,11}|$ versus $\lambda_4$ in Fig.~\ref{Fig2} (a), (b), (c) respectively, where the solid, dashed, dotted lines denote the results for $\Lambda=10,\;12,\;14\;{\rm TeV}$ respectively. The gray areas denote the range $94\;{\rm GeV}<m_{h_1}<96\;{\rm GeV}$ for Fig.~\ref{Fig2} (a) and $124\;{\rm GeV}<m_{h_2}<126\;{\rm GeV}$ for Fig.~\ref{Fig2} (b). It can be observed from Fig.~\ref{Fig2} that $m_{h_1}$ increases with the increasing of $\lambda_4$, while $m_{h_2}$ decreases with the increasing of $\lambda_4$. Both $m_{h_1}$ and $m_{h_2}$ increase as $\Lambda$ increasing. Fig.~\ref{Fig2} shows that in our chosen parameter space, $\lambda_4$ is around $0.00002$ when $m_{h_1}\approx 95\;{\rm GeV}$ and $m_{h_2}\approx 125\;{\rm GeV}$.

Based on the preceding analysis, it is evident that both the $95\;{\rm GeV}$ lightest Higgs boson and the $125\;{\rm GeV}$ SM-like Higgs boson are attainable in the model, and $\lambda_1$, $\lambda_2$, $\lambda_4$, $\Lambda$ collectively determine the two light Higgs masses. Subsequently, we investigate the potential of the $95\;{\rm GeV}$ Higgs boson within the flavor-dependent $U_1(x)$ model to account for the observed excesses in diphoton and ditau events.
\begin{figure}
\setlength{\unitlength}{1mm}
\centering
\includegraphics[width=2.1in]{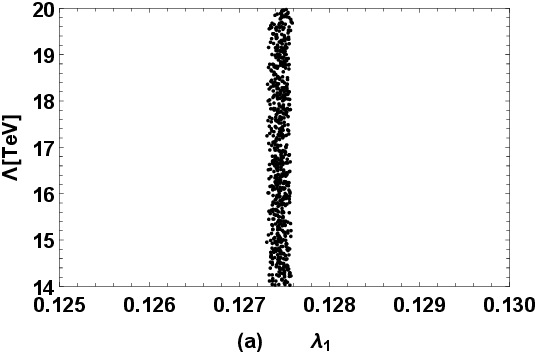}
\vspace{0.5cm}
\includegraphics[width=2.1in]{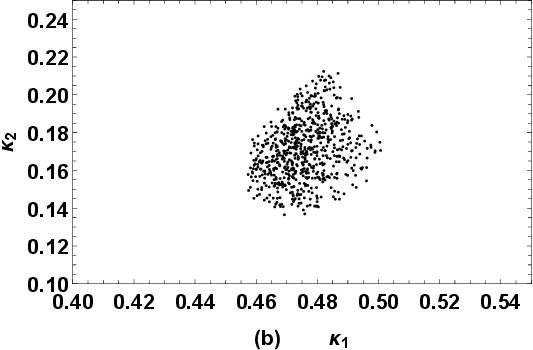}
\vspace{0.5cm}
\includegraphics[width=2.1in]{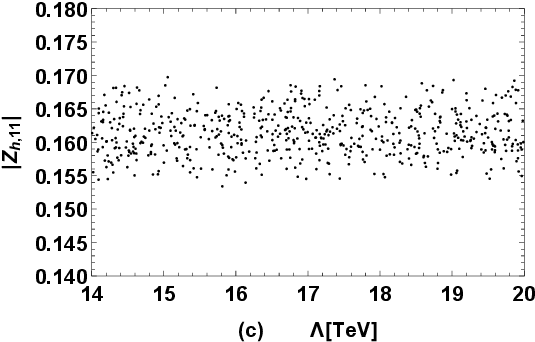}
\vspace{0cm}
\caption[]{Keeping $m_{h_1}=95\;{\rm GeV}$, $m_{h_2}=125\;{\rm GeV}$, the measured $125$ GeV Higgs signals in the range shown in Eq.~(\ref{125S}), $0.21<\mu(h_{95})_{\gamma\gamma}<0.52$, $0.7<\mu(h_{95})_{\tau\tau}<1.7$, and scanning the parameter space in Eq.~(\ref{eq28}), the allowed ranges of $\lambda_1-\Lambda$ (a), $\kappa_1-\kappa_2$ (b) are plotted. In the allowed ranges shown in (a) and (b), the Higgs mixing parameter $|Z_{h,11}|$ versus $\Lambda$ is plotted in (c).}
\label{Fig3}
\end{figure}
To comprehensively understand the collective influences of $\lambda_1$, $\lambda_2$, $\lambda_4$, $\Lambda$, $\kappa_1$ and $\kappa_2$ on the signal strength of diphoton and ditau events, we scan the following parameter space
\begin{eqnarray}
&&\lambda_1=(0.0001,0.2),\;\Lambda=(14,20)\;{\rm TeV},\;\kappa_1=(0.0001,1)\;,\;\kappa_2=(0.0001,1),\label{eq28}
\end{eqnarray}
where $\lambda_2$, $\lambda_4$ are determined by
\begin{eqnarray}
&&\lambda_2=\frac{(m_{h_1}^2+m_{h_2}^2-2\lambda_1 v^2)}{2\Lambda^2},\;\lambda_4=\frac{\sqrt{(m_{h_1}^2-2\lambda_1 v^2)(2\lambda_1\;v^2-m_{h_2}^2)}}{v\;\Lambda},\label{aL}
\end{eqnarray}
with $m_{h_1}=95\;{\rm GeV}$ and $m_{h_2}=125\;{\rm GeV}$. Keeping the measured $125$ GeV Higgs signals in the range~\cite{PDG}
\begin{eqnarray}
&&\mu_{WW^*}=1.00\pm0.08,\;\;\mu_{ZZ^*}=1.02\pm0.08,\;\;\mu_{\gamma\gamma^*}=1.10\pm0.07,\nonumber\\
&&\mu_{c \bar{c}}=8\pm22,\;\;\mu_{b \bar{b}}=0.99\pm0.12,\;\;\mu_{\mu \bar\mu}=1.21\pm0.35,\mu_{\tau \bar\tau}=0.99\pm0.12,\label{125S}
\end{eqnarray}
and the two CMS excesses at about $95$ GeV in the ranges $0.21<\mu(h_{95})_{\gamma\gamma}<0.52$, $0.7<\mu(h_{95})_{\tau\tau}<1.7$, we plot the allowed ranges of $\lambda_1-\Lambda$, $\kappa_1-\kappa_2$ in Fig.~\ref{Fig3} (a), Fig.~\ref{Fig3} (b) respectively. In the allowed ranges shown in Fig.~\ref{Fig3} (a) and Fig.~\ref{Fig3} (b), the Higgs mixing parameter $|Z_{h,11}|$ versus $\Lambda$ is plotted in Fig.~\ref{Fig3} (c).

It is obvious from Fig.~\ref{Fig3} that the flavor-dependent $U(1)_X$ model can account well for both the diphoton and ditau excesses at about $95\;{\rm GeV}$. Fig.~\ref{Fig3} (a) shows that the parameter $\Lambda$ is not strictly limited by the diphoton and ditau excesses, while $\lambda_1$ is strictly limited. Because $\Lambda$ affects the 95~GeV Higgs signals and 125~GeV Higgs signals mainly through the mixing effects of $S_1$ and $S_2$ as shown in Eq.~(\ref{Zh}). It is obvious that the dependence of $A$ on $\Lambda$ by the terms $\lambda_4\;\Lambda$ and $\lambda_2\;\Lambda^2$, and both the two terms depend on $\lambda_1$ for fixed $m_{h_1}$, $m_{h_2}$ as shown in Eq.~(\ref{aL}). Hence, the excesses are satisfied quite independent of the value of $\Lambda$, while $\lambda_1$ is strictly limited. In addition, Eq.~(\ref{amh}) shows that the SM-like Higgs mass $m_{h_2}$ mainly depends on $\lambda_1$ which also leads to the strict constraint on $\lambda_1$. $\kappa_1$ and $\kappa_2$ are also limited strictly as shown in Fig.~\ref{Fig3} (b), by all the constrains taken into consideration. The diphoton and ditau excesses at about $95\;{\rm GeV}$ prefer the ranges $0.128\gtrsim\lambda_1\gtrsim0.127$, $0.52\gtrsim\kappa_1\gtrsim0.45$ and $0.22\gtrsim\kappa_2\gtrsim0.12$. Fig.~\ref{Fig3} (c) shows that $|Z_{h,11}|$ is located in the range $0.15\lesssim|Z_{h,11}|\lesssim0.17$, because the $95$ GeV state is dominated by the extra singlet scalar in the model and the Higgs squared mass matrix in Eq.~(\ref{eq8}) is written on the basis $(S_1,\;S_2,\;S_3)$. Finally, provided that the excesses persist, the precision measurement of the 125 GeV Higgs couplings has potential to detect this scenario, because the introducing of a light Higgs boson below 125 GeV changes the coupling properties of 125 GeV Higgs.

\section{Summary\label{sec4}}

Considering the diphoton and ditau excesses at a mass about $95\;{\rm GeV}$ in the CMS data, we focus on explaining these two excesses in flavor-dependent $U(1)_X$ model, because the model can produces a $95\;{\rm GeV}$ light Higgs naturally by introducing the specific singlet state. And this new scalar state interacts with the quark sector directly to explain the flavor numbers of the fermion sector, which provides great potential to explain the two excesses. Considering the specific Higgs state as the lightest Higgs boson, it is found that the specific parameters $\lambda_1$, $\lambda_2$, $\lambda_4$ and $\Lambda$ in the model affects the theoretical predictions on the two light Higgs boson masses significantly. Taking the lightest Higgs boson mass at about $95\;{\rm GeV}$ and the next-to-lightest Higgs boson mass at about $125\;{\rm GeV}$, the numerical results show that the specific Higgs state in flavor-dependent $U(1)_X$ model can accounts for the observed CMS diphoton and ditau excesses at about $95\;{\rm GeV}$ well. In addition, fitting $\mu(h_{95})_{\gamma\gamma}$, $\mu(h_{95})_{\tau\tau}$ in the CMS $1\sigma$ interval will impose stringent constraints on the parameter space of the flavor-dependent $U(1)_X$ model, and the fact can be seen explicitly in Fig.~\ref{Fig3}.

\begin{acknowledgments}

The work has been supported by the National Natural Science Foundation of China (NNSFC) with Grants No. 12075074, No. 12235008, Hebei Natural Science Foundation for Distinguished Young Scholars with Grant No. A2022201017, Natural Science Foundation of Guangxi Autonomous Region with Grant No. 2022GXNSFDA035068, and the youth top-notch talent support program of the Hebei Province.

\end{acknowledgments}

\end{CJK*}


\begin{thebibliography}{99}
\bibitem{CMS:2012qbp}S.~Chatrchyan \textit{et al.} [CMS], Phys. Lett. B \textbf{716}, 30-61 (2012).
\bibitem{ATLAS:2012yve}G.~Aad \textit{et al.} [ATLAS], Phys. Lett. B \textbf{716}, 1-29 (2012).
\bibitem{CMS:2022dwd}A.~Tumasyan \textit{et al.} [CMS], Nature \textbf{607}, no.7917, 60-68 (2022).
\bibitem{ATLAS:2022vkf}[ATLAS], Nature \textbf{607}, no.7917, 52-59 (2022) [erratum: Nature \textbf{612}, no.7941, E24 (2022)].

\bibitem{CMS:2015ocq}[CMS], CMS-PAS-HIG-14-037.
\bibitem{CMS:2018cyk}A.~M.~Sirunyan \textit{et al.} [CMS], Phys. Lett. B \textbf{793}, 320-347 (2019).
\bibitem{ATLAS:2023CA}C. Arcangeletti. on behalf of ATLAS collaboration, LHC Seminar, 7th of June, 2023.

\bibitem{ATLAS:2018xad}[ATLAS], ATLAS-CONF-2018-025.
\bibitem{CMS:2022goy}A.~Tumasyan \textit{et al.} [CMS], JHEP \textbf{07}, 073 (2023).
\bibitem{CMS:2023}S. Gascon-Shotkin, CMS, Talk at MoriondEW: Searches for additional Higgs bosons at low mass, 2023.


\bibitem{Moretti:2006sv}S.~Moretti and S.~Munir, Eur. Phys. J. C \textbf{47}, 791-803 (2006).
\bibitem{Ellwanger:2010nf}U.~Ellwanger, Phys. Lett. B \textbf{698}, 293-296 (2011).
\bibitem{Cao:2011pg}J.~Cao, Z.~Heng, T.~Liu and J.~M.~Yang, Phys. Lett. B \textbf{703}, 462-468 (2011).
\bibitem{AlbornozVasquez:2012foo}D.~Albornoz Vasquez, G.~Belanger, C.~Boehm, J.~Da Silva, P.~Richardson and C.~Wymant, Phys. Rev. D \textbf{86}, 035023 (2012).
\bibitem{Ellwanger:2012ke}U.~Ellwanger and C.~Hugonie,n Adv. High Energy Phys. \textbf{2012}, 625389 (2012).
\bibitem{Boudjema:2012in}F.~Boudjema and G.~D.~La Rochelle, Phys. Rev. D \textbf{86}, 115007 (2012).
\bibitem{Schmidt-Hoberg:2012dba}K.~Schmidt-Hoberg and F.~Staub, JHEP \textbf{10}, 195 (2012).
\bibitem{Badziak:2013bda}M.~Badziak, M.~Olechowski and S.~Pokorski, JHEP \textbf{06}, 043 (2013).
\bibitem{Badziak:2013gla}M.~Badziak, M.~Olechowski and S.~Pokorski, PoS \textbf{EPS-HEP2013}, 257 (2013).
\bibitem{Barbieri:2013nka}R.~Barbieri, D.~Buttazzo, K.~Kannike, F.~Sala and A.~Tesi, Phys. Rev. D \textbf{88}, 055011 (2013).
\bibitem{Fan:2013gjf}J.~W.~Fan, J.~Q.~Tao, Y.~Q.~Shen, G.~M.~Chen, H.~S.~Chen, S.~Gascon-Shotkin, M.~Lethuillier, L.~Sgandurra and P.~Soulet, Chin. Phys. C \textbf{38}, 073101 (2014).
\bibitem{Potter:2015wsa}C.~T.~Potter, Eur. Phys. J. C \textbf{76}, no.1, 44 (2016).
\bibitem{Ellwanger:2015uaz}U.~Ellwanger and M.~Rodriguez-Vazquez, JHEP \textbf{02}, 096 (2016).
\bibitem{Cao:2016uwt}J.~Cao, X.~Guo, Y.~He, P.~Wu and Y.~Zhang, Phys. Rev. D \textbf{95}, no.11, 116001 (2017).
\bibitem{Cao:2019ofo}J.~Cao, X.~Jia, Y.~Yue, H.~Zhou and P.~Zhu, Phys. Rev. D \textbf{101}, no.5, 055008 (2020).




\bibitem{Biekotter:2019kde}T.~Biek\"otter, M.~Chakraborti and S.~Heinemeyer, Eur. Phys. J. C \textbf{80}, no.1, 2 (2020).
\bibitem{Biekotter:2021ovi}T.~Biek\"otter and M.~O.~Olea-Romacho, JHEP \textbf{10}, 215 (2021).
\bibitem{Biekotter:2021qbc}T.~Biek\"otter, A.~Grohsjean, S.~Heinemeyer, C.~Schwanenberger and G.~Weiglein,Eur. Phys. J. C \textbf{82}, no.2, 178 (2022).
\bibitem{Heinemeyer:2021msz}S.~Heinemeyer, C.~Li, F.~Lika, G.~Moortgat-Pick and S.~Paasch, Phys. Rev. D \textbf{106}, no.7, 075003 (2022).
\bibitem{Biekotter:2022jyr}T.~Biek\"otter, S.~Heinemeyer and G.~Weiglein, JHEP \textbf{08}, 201 (2022).
\bibitem{Biekotter:2022abc}T.~Biek\"otter, S.~Heinemeyer and G.~Weiglein, Eur. Phys. J. C \textbf{83}, no.5, 450 (2023).
\bibitem{Biekotter:2023jld}T.~Biek\"otter, S.~Heinemeyer and G.~Weiglein, [arXiv:2303.12018 [hep-ph]].
\bibitem{Azevedo:2023zkg}D.~Azevedo, T.~Biek\"otter and P.~M.~Ferreira, [arXiv:2305.19716 [hep-ph]].
\bibitem{Aguilar-Saavedra:2023vpd}J.~A.~Aguilar-Saavedra, H.~B.~C\^amara, F.~R.~Joaquim and J.~F.~Seabra, [arXiv:2307.03768 [hep-ph]].


\bibitem{Sachdeva:2019hvk}D.~Sachdeva and S.~Sadhukhan, Phys. Rev. D \textbf{101}, no.5, 055045 (2020).

\bibitem{Biekotter:2017xmf}T.~Biek\"otter, S.~Heinemeyer and C.~Mu\~noz, Eur. Phys. J. C \textbf{78}, no.6, 504 (2018).
\bibitem{Biekotter:2019gtq}T.~Biek\"otter, S.~Heinemeyer and C.~Mu\~noz, Eur. Phys. J. C \textbf{79}, no.8, 667 (2019).

\bibitem{LEP:2003ing}R.~Barate \textit{et al.} [LEP Working Group for Higgs boson searches, ALEPH, DELPHI, L3 and OPAL], Phys. Lett. B \textbf{565}, 61-75 (2003).


\bibitem{Ashanujjaman:2023etj}S.~Ashanujjaman, S.~Banik, G.~Coloretti, A.~Crivellin, B.~Mellado and A.~T.~Mulaudzi, [arXiv:2306.15722 [hep-ph]].

\bibitem{Azatov:2012bz}A.~Azatov, R.~Contino and J.~Galloway, JHEP \textbf{04}, 127 (2012).
\bibitem{Heinemeyer:2018wzl}S.~Heinemeyer and T.~Stefaniak, PoS \textbf{CHARGED2018}, 016 (2019).



\bibitem{VanLoi:2023utt}D.~Van Loi and P.~Van Dong, Eur. Phys. J. C \textbf{83}, no.11, 1048 (2023).

\bibitem{CMS:2022arx}[CMS], CMS-PAS-EXO-21-018.
\bibitem{Iguro:2022dok}S.~Iguro, T.~Kitahara and Y.~Omura, Eur. Phys. J. C \textbf{82}, no.11, 1053 (2022).

\bibitem{Chivukula:1987py}R.~S.~Chivukula and H.~Georgi, Phys. Lett. B \textbf{188}, 99-104 (1987).
\bibitem{Hall:1990ac}L.~J.~Hall and L.~Randall, Phys. Rev. Lett. \textbf{65}, 2939-2942 (1990).
\bibitem{Buras:2000dm}A.~J.~Buras, P.~Gambino, M.~Gorbahn, S.~Jager and L.~Silvestrini, Phys. Lett. B \textbf{500}, 161-167 (2001).
\bibitem{DAmbrosio:2002vsn}G.~D'Ambrosio, G.~F.~Giudice, G.~Isidori and A.~Strumia, Nucl. Phys. B \textbf{645}, 155-187 (2002).
\bibitem{Isidori:2010kg}G.~Isidori, Y.~Nir and G.~Perez, Ann. Rev. Nucl. Part. Sci. \textbf{60}, 355 (2010).


\bibitem{Djouadi:1997yw}A.~Djouadi, J.~Kalinowski and M.~Spira, Comput. Phys. Commun. \textbf{108}, 56-74 (1998).
\bibitem{Arbey} A. Arbey, A. Deandrea, F. Mahmoudi, and A. Tarhini, Anomaly mediated supersymmetric models and Higgs data from the LHC, Phys. Rev. D 87, 115020 (2013), arXiv:1304.0381 [hep-ph].
\bibitem{PDG}R.~L.~Workman \textit{et al.} [Particle Data Group], PTEP \textbf{2022}, 083C01 (2022).








\end{thebibliography}
\end{document}